\documentstyle[preprint,eqsecnum,epsfig,aps,floats,tighten]{revtex}
\begin{document}
\lefthyphenmin=2
\righthyphenmin=3
%%%%%%%%%%%%%%%%%%%%%%%%%%	DEFINITIONS	%%%%%%%%%%%%%%%%%%%%%%%%%%%%%
\newcommand{\rstev}{\mbox{$\rs = \T{1.8}$}}
\newcommand{\T}[1]{\mbox{#1 TeV}}
\newcommand{\rs}{\mbox{$\sqrt{s}$}}

\newcommand{\XX}{\mbox{$\, \times \,$}}
\newcommand{\DP}{\mbox{$\Delta\phi$}}
\newcommand{\DE}{\mbox{$\Delta\eta$}}
\newcommand{\GEANT}{\footnotesize\rm GEANT \normalsize}
\newcommand{\ISAJET}{\footnotesize\rm ISAJET \normalsize}

\def\munot{$\mu_0 = (P_T^2 + M_b^2)^{1/2}$}
\def\mujot{$\mu_0 = (E_T^2 + M_b^2)^{1/2}$}

\def\pbarp{$\overline{p}p $}            %pbarp
\def\ipb{pb$^{-1}$}                     %inverse picobarns
\def\pt{$p_T$}                          %pT
\def\D0{D\O}                            %D0
\def\ptrel{$p_T^{\rm rel}$}		%PtRel
\def\into{\rightarrow}                

% A useful Journal macro
\def\Journal#1#2#3#4{{#1}{\bf #2}, #3 (#4)}
% Some useful journal names
\def\NIMA{Nucl. Instrum. Methods Phys. Res. A }
\def\NPB{{Nucl. Phys.} {\bf B}}
\def\PLB{Phys. Lett. B }
\def\PRL{Phys. Rev. Lett. }
\def\PRD{Phys. Rev. D }
\def\ZPC{Z. Phys. C }
\def\EPJC{Eur. Phys. J. C }

%%%%%%%%%%%%%%%%%%%%%%%%%%	TITLE/AUTHORS	%%%%%%%%%%%%%%%%%%%%%%%%%%%%%
\boldmath
\title{Cross Section for $b$ Jet Production 
in \pbarp\ Collisions at \rstev}
\unboldmath
%%%%%%%%%%%%%
%%\input{../list_of_authors.tex
% LIST_OF_AUTHORS.TEX                 8/10/00            
\author{                                                                      
%% names begin here                                                           
B.~Abbott,$^{50}$                                                             
M.~Abolins,$^{47}$                                                            
V.~Abramov,$^{23}$                                                            
B.S.~Acharya,$^{15}$                                                          
D.L.~Adams,$^{57}$                                                            
M.~Adams,$^{34}$                                                              
G.A.~Alves,$^{2}$                                                             
N.~Amos,$^{46}$                                                               
E.W.~Anderson,$^{39}$                                                         
M.M.~Baarmand,$^{52}$
% ...
\\                                                                            
\vskip 0.30cm                                                                 
\centerline{(D\O\ Collaboration)}                                             
\vskip 0.30cm                                                                 
}                                                                             
\address{                                                                     
\centerline{$^{1}$Universidad de Buenos Aires, Buenos Aires, Argentina}       
\centerline{$^{2}$LAFEX, Centro Brasileiro de Pesquisas F{\'\i}sicas,         
                  Rio de Janeiro, Brazil}                                     
\centerline{$^{3}$Universidade do Estado do Rio de Janeiro,                   
                  Rio de Janeiro, Brazil}                                     
\centerline{$^{4}$Institute of High Energy Physics, Beijing,                  
                  People's Republic of China}                                 
\centerline{$^{5}$Universidad de los Andes, Bogot\'{a}, Colombia}             
\centerline{$^{6}$Charles University, Prague, Czech Republic}                 
\centerline{$^{7}$Institute of Physics, Academy of Sciences, Prague,          
                  Czech Republic}                                             
\centerline{$^{8}$Universidad San Francisco de Quito, Quito, Ecuador}         
\centerline{$^{9}$Institut des Sciences Nucl\'eaires, IN2P3-CNRS,             
                  Universite de Grenoble 1, Grenoble, France}                 
\centerline{$^{10}$CPPM, IN2P3-CNRS, Universit\'e de la M\'editerran\'ee,     
                  Marseille, France}                                          
\centerline{$^{11}$LPNHE, Universit\'es Paris VI and VII, IN2P3-CNRS,         
                  Paris, France}                                              
\centerline{$^{12}$DAPNIA/Service de Physique des Particules, CEA, Saclay,    
                  France}                                                     
\centerline{$^{13}$Panjab University, Chandigarh, India}                      
\centerline{$^{14}$Delhi University, Delhi, India}                            
\centerline{$^{15}$Tata Institute of Fundamental Research, Mumbai, India}     
\centerline{$^{16}$Seoul National University, Seoul, Korea}                   
\centerline{$^{17}$CINVESTAV, Mexico City, Mexico}                            
\centerline{$^{18}$FOM-Institute NIKHEF and University of                     
                  Amsterdam/NIKHEF, Amsterdam, The Netherlands}               
\centerline{$^{19}$University of Nijmegen/NIKHEF, Nijmegen, The               
                  Netherlands}                                                
\centerline{$^{20}$Institute of Nuclear Physics, Krak\'ow, Poland}            
\centerline{$^{21}$Institute for Theoretical and Experimental Physics,        
                   Moscow, Russia}                                            
\centerline{$^{22}$Moscow State University, Moscow, Russia}                   
\centerline{$^{23}$Institute for High Energy Physics, Protvino, Russia}       
\centerline{$^{24}$Lancaster University, Lancaster, United Kingdom}           
\centerline{$^{25}$University of Arizona, Tucson, Arizona 85721}              
\centerline{$^{26}$Lawrence Berkeley National Laboratory and University of    
                  California, Berkeley, California 94720}                     
\centerline{$^{27}$University of California, Davis, California 95616}         
\centerline{$^{28}$California State University, Fresno, California 93740}     
\centerline{$^{29}$University of California, Irvine, California 92697}        
\centerline{$^{30}$University of California, Riverside, California 92521}     
\centerline{$^{31}$Florida State University, Tallahassee, Florida 32306}      
\centerline{$^{32}$University of Hawaii, Honolulu, Hawaii 96822}              
\centerline{$^{33}$Fermi National Accelerator Laboratory, Batavia,            
                   Illinois 60510}                                            
\centerline{$^{34}$University of Illinois at Chicago, Chicago,                
                   Illinois 60607}                                            
\centerline{$^{35}$Northern Illinois University, DeKalb, Illinois 60115}      
\centerline{$^{36}$Northwestern University, Evanston, Illinois 60208}         
\centerline{$^{37}$Indiana University, Bloomington, Indiana 47405}            
\centerline{$^{38}$University of Notre Dame, Notre Dame, Indiana 46556}       
\centerline{$^{39}$Iowa State University, Ames, Iowa 50011}                   
\centerline{$^{40}$University of Kansas, Lawrence, Kansas 66045}              
\centerline{$^{41}$Kansas State University, Manhattan, Kansas 66506}          
\centerline{$^{42}$Louisiana Tech University, Ruston, Louisiana 71272}        
\centerline{$^{43}$University of Maryland, College Park, Maryland 20742}      
\centerline{$^{44}$Boston University, Boston, Massachusetts 02215}            
\centerline{$^{45}$Northeastern University, Boston, Massachusetts 02115}      
\centerline{$^{46}$University of Michigan, Ann Arbor, Michigan 48109}         
\centerline{$^{47}$Michigan State University, East Lansing, Michigan 48824}   
\centerline{$^{48}$University of Nebraska, Lincoln, Nebraska 68588}           
\centerline{$^{49}$Columbia University, New York, New York 10027}             
\centerline{$^{50}$New York University, New York, New York 10003}             
\centerline{$^{51}$University of Rochester, Rochester, New York 14627}        
\centerline{$^{52}$State University of New York, Stony Brook,                 
                   New York 11794}                                            
\centerline{$^{53}$Brookhaven National Laboratory, Upton, New York 11973}     
\centerline{$^{54}$Langston University, Langston, Oklahoma 73050}             
\centerline{$^{55}$University of Oklahoma, Norman, Oklahoma 73019}            
\centerline{$^{56}$Brown University, Providence, Rhode Island 02912}          
\centerline{$^{57}$University of Texas, Arlington, Texas 76019}               
\centerline{$^{58}$Texas A\&M University, College Station, Texas 77843}       
\centerline{$^{59}$Rice University, Houston, Texas 77005}                     
\centerline{$^{60}$University of Washington, Seattle, Washington 98195}       
}                                                                             
%end                                                                          
%%%%%%%%%%%%%
\date{\today}
\maketitle
%%%%%%%%%%%%%%%%%%%%%%%%%%%%%%%%%%%%%%%%%%%%%%%%%%%%%%%%%%%%%%%%%%%%%%%%%%%%%
\begin{abstract}
Bottom quark production in \pbarp\ collisions at $\sqrt{s}=1.8$ TeV 
is studied with 5 \ipb\ of data collected in 1995 by
the D\O\ detector at the Fermilab Tevatron Collider.
The differential production cross section for $b$ jets
in the central rapidity region ($|y^b|<1$)
as a function of jet transverse energy is extracted from a muon-tagged jet
sample.  Within experimental and theoretical uncertainties, D\O\
results are found to be higher than, but compatible with,
next-to-leading-order QCD  predictions.
\end{abstract}
\pacs{PACS numbers (empty)}

%%\twocolumn
%%%%%%%%%%%%%%%%%%%%%%%%%%%%%%%%%%%%%%%%%%%%%%%%%%%%%%%%%%%%%%%%%%%%%%%%%%%%%
%\section*{Introduction - Status}
Measurements of the bottom-quark production cross section at \pbarp\ 
colliders provide an important quantitative test of quantum
chromodynamics (QCD). The mass of the $b$ quark is considered large enough 
($m_b\gg\Lambda_{\rm QCD}$) to justify perturbative expansions in the 
strong coupling constant $\alpha_s$. Consequently, data on 
$b$ quark production are expected to be 
adequately described by calculations 
at next-to-leading order (NLO) in $\alpha_s$\cite{bib:nde,bib:bee,bib:mnr}.

Past measurements of inclusive $b$ quark production in the central
rapidity region, at center-of-mass energies $\sqrt{s}=0.63$ TeV
\cite{bib:ua1,bib:stich} and $\sqrt{s}=1.8$ TeV 
\cite{bib:cdfsgmu,bib:d0sgmu,bib:d0psi,bib:d0corr},
indicate a general agreement in shape with the calculated 
transverse momentum (\pt)
spectrum, but are systematically higher than the NLO QCD
predictions\cite{bib:nde,bib:bee,bib:mnr} by roughly a factor of 2.5.
More recently, a measurement using muons at high rapidity 
\cite{bib:samus} ($2.4<|y^{\mu}|<3.2$) to tag $b$ quarks,
indicates an even larger excess of 
observed cross section over theory. 
The measured differential production cross section for $B$ mesons
\cite{bib:cdfmes} is similarly higher than the NLO QCD prediction.
Calculations of higher order corrections \cite{bib:tung,bib:grecco} 
have shown that 
additional enhancements to the cross section beyond NLO are likely.
However, the expected enhancements fall short of accounting for the observed
discrepancy between theory and data. 
This longstanding mismatch has
motivated continuous theoretical and experimental effort dedicated to
reduce uncertainties in general, and to broaden the scope of
observable quantities \cite{bib:d0corr,bib:cdfcorr}. 

Previous studies of $b$ quark production by the \D0\ collaboration have
exploited the kinematic relationship between $b$ quarks and daughter 
(semileptonic decay) muons to extract integrated 
$b$ quark production rates as a function of  $p_T^b$ threshold 
\cite{bib:d0sgmu},  and have examined 
azimuthal correlations between the $b$ and $\overline{b}$
in pair production \cite{bib:d0corr}.

The present study is a complementary measurement of $b$ production, 
based primarily on calorimetry, 
with the main focus being on $b$ jets rather than $b$ quarks. 
$b$ jets are defined as hadronic jets carrying $b$ flavor. 
As opposed to quarks, jets are directly observable and therefore reduce 
model dependence when comparing experimental data with theory.
This measurement is in direct correspondence with a NLO QCD calculation
\cite{bib:fmnr} that highlights the advantages of considering $b$ jets rather 
than open $b$ quarks. For instance, large logarithms that appear at all orders 
in the open quark calculation (due to hard collinear gluons) are avoided
when all fragmentation modes are integrated.

The differential production cross section of $b$ jets as a function
of the jet transverse energy ($E_T$) has been measured using data collected 
during 1995 with the \D0\ detector at the Fermilab Tevatron Collider. 
The data correspond to an integrated luminosity of $(5.2\pm0.3)$ \ipb.
The analysis exploits the semileptonic decays of $b$ hadrons which
result in a muon associated with a jet. 
An inclusive sample of muon-tagged jets is selected from a trigger requiring 
a jet and a muon, and the $b$ jet component of the inclusive sample is 
extracted, based on the properties expected of the associated muon and jet.

%%%%%%%%%%%%%%%%%%%%%%%%%%%%%%%%%%%%%%%%%%%%%%%%%%%%%%%%%%%%%%%%%%%%%%%%%%%%%
%\section*{Detector}
The D\O\ detector and trigger system are described
elsewhere\cite{bib:d0nim}.
Jet detection utilizes primarily the uranium-liquid argon 
calorimeters, which have full coverage for pseudorapidity 
\cite{bib:eta} $|\eta| \leq 4$, and are segmented into towers of 
$\DE \XX \DP = 0.1 \XX 0.1$, where $\phi$ is the azimuthal angle.
The relative energy resolution for jets is approximately 80\%
/$\sqrt{E({\rm GeV})}$.
The central muon system consists of three layers of proportional drift 
tubes and a magnetized iron toroid located between the first two layers. 
The muon detectors
provide a measurement of the muon momentum with a resolution parameterized by
$\delta(1/p) = 0.18(p-2)/p^2 \oplus 0.003$, with $p$ in GeV/$c$. 
The total thickness of the calorimeter and toroid in the
central region varies from 13 to 15 interaction lengths, which reduces the
hadronic punchthrough for the muon system to less than 0.5\% 
of muons from all sources.

%%%%%%%%%%%%%%%%%%%%%%%%%%%%%%%%%%%%%%%%%%%%%%%%%%%%%%%%%%%%%%%%%%%%%%%%%%%%%
%\section*{Event Selection - Trig&Reco&Offline}
 Initial event selection used a trigger requiring
(i) $E_T >$ 10 GeV in at least one
calorimeter trigger tile of $\DE \XX \DP = 0.8 \XX 1.6$,
(ii) at least one muon candidate with $p_T >$ 3 GeV/$c$, and
(iii) a single \pbarp\ interaction per beam crossing 
as signaled by the trigger scintillation hodoscopes.

Jet candidates were reconstructed offline with an iterative fixed-cone 
algorithm (cone radius of 0.7 in $\eta$-$\phi$ space)
and then subjected to quality selection criteria to eliminate background 
from isolated noisy calorimeter cells and accelerator beam losses which 
may mimic jets. Accepted jets were required to have 
$E_T>25$ GeV, $|\eta|<0.6$,
and an associated muon within the reconstruction cone.

Offline muon selection required the triggered track to originate from the 
reconstructed event vertex, with $p_T^{\mu}>6$ GeV/$c$ and $|\eta^{\mu}|<0.8$. 
Muon candidates in the azimuth region $80^{\circ}<\phi^{\mu}<110^{\circ}$ 
were excluded due to poor chamber efficiencies near where
the Main Ring beam pipe passes through D\O.

Background events from cosmic ray muons were eliminated by requiring 
a 10 ns wide time coincidence of muon tracks with beam crossings.
About 30,000 events were selected, of which less than 0.5\%
have either a second muon-tagged jet or a double-tagged one. 
This is in agreement with expectations from muon cuts and acceptance,
branching fractions, and flavor content of the sample.

%%%%%%%%%%%%%%%%%%%%%%%%%%%%%%%%%%%%%%%%%%%%%%%%%%%%%%%%%%%%%%%%%%%%%%%%%%%%
%\section*{Definition of XSecn}
The differential $b$ jet cross section as a function of jet $E_T$ 
is extracted from the inclusive tagged jet sample as
\begin{equation}
\frac{d\sigma}{dE_T^{b\rm Jet}}  =
\frac{U(E_T) F_b(E_T)}
	{2B{\cal L}_{\rm int}~\epsilon^{\mu J}(p_T^{\mu},E_T) A(E_T)}
\frac{\Delta N^{\mu J}}{\Delta E_{T}}
\label{eqn:bxsec}
\end{equation}

where 
${\Delta N^{\mu J}}/{\Delta E_{T}}$ represents the inclusive tagged jet 
counts per transverse energy bin-width,
$F_b(E_T)$ is the fraction of tagged jets 
containing muons from $b$ quark decays,
$\epsilon^{\mu J}(p_T^{\mu},E_T)$ is the event detection efficiency 
of a $\mu +$jet pair,
$U(E_T)$ corrects for spectrum smearing from jet energy resolution,
$A(E_T)$ corrects for the muon tagging acceptance,
${\cal L}_{\rm int}$ is the integrated luminosity of the data sample, and
$B = 0.108\pm0.005$ is the branching fraction
for inclusive decays $b \into \mu + X$ \cite{bib:pdg}.
Tagging muons of both charges are counted, and  
the number of events is divided by two. The cross
section is determined for the $(b+\bar{b})/2$ combination.
In Eq.~\ref{eqn:bxsec} and throughout this paper, $E_T$ represents the
calorimeter-only component of the jet transverse energy: it excludes the
tagging muon and associated neutrino.
Once the correction for the lepton energy is included
(see below), then $E_T^{b\rm Jet}$ measures the complete $b$ jet.

%%%%%%%%%%%%%%%%%%%%%%%%%%%%%%%%%%%%%%%%%%%%%%%%%%%%%%%%%%%%%%%%%%%%%%%%%%%%%
%\section*{Data Correction; Effs, Resoln - Trig&Reco&Offline}
The trigger and offline reconstruction efficiencies were determined from 
Monte Carlo events, and crosschecked with appropriate control samples of
collider data. Monte Carlo events were generated with \ISAJET 
\cite{bib:isajet}, followed by a \GEANT \cite{bib:geant} simulation of
the D\O\ detector response, trigger simulation, and reconstruction.
The overall muon geometric acceptance multiplied by efficiency 
increases from 35\% 
at $p_T^{\mu}=$ 6 GeV/$c$ to a plateau of 45\% above 10 GeV/$c$.
The overall jet efficiency increases from 85\% 
at $E_T=$ 25 GeV to a plateau of 97\% above 45 GeV.

The transverse energy of each jet was corrected for the 
underlying event, additional interactions, noise from uranium decay,
the fraction of particle energy showered outside the jet cone, detector
uniformity, and detector hadronic response \cite{bib:etsc}.  
The steeply falling $E_T$ spectrum is distorted by jet energy resolution,
and was corrected as described in \cite{bib:d0incj}.

%%%%%%%%%%%%%%%%%%%%%%%%%%%%%%%%%%%%%%%%%%%%%%%%%%%%%%%%%%%%%%%%%%%%%%%%%%%%
%\section*{Signal and Background Determination}
\begin{figure}
\vspace{-1.0cm}
\begin{center}
\mbox{\epsfig{figure=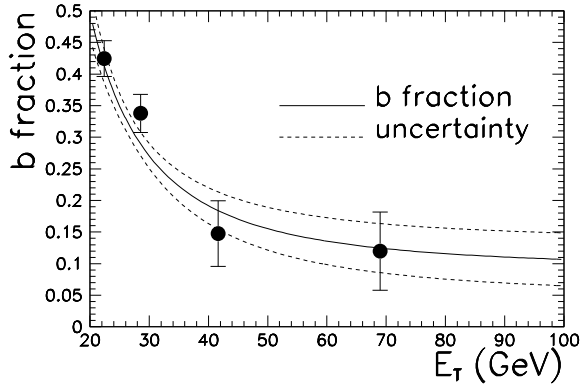,height=6cm,width=8cm}}
\end{center}
\vspace{-0.4cm}
\caption{Fraction of muon-tagged jets which come from b decays.
$E_T$ is the calorimeter-only 
component (see text) of the $b$ jet transverse energy.}
\label{fig:ptrel}
\end{figure}

In addition to $b$ production, muon-tagged jets can also arise from 
semileptonic decays of $c$ quarks, and in-flight decays of $\pi$ or $K$ 
mesons. Muons from Drell-Yan or on-shell $W/Z$ boson production are not 
expected to have associated jets, and Monte Carlo estimates confirm a 
negligible contribution.
The background from light flavors is suppressed using
the transverse momentum of the muon relative to the associated 
jet axis (\ptrel) as discriminator. 
The higher mass of the decaying $b$ quark generates a rather different 
\ptrel\ distribution than obtained from quark decays of lighter flavor.

Individual \ptrel\ distributions for $b$, $c$, and $\pi/K$ decays were 
modeled using the \ISAJET \cite{bib:isajet} Monte Carlo,
with individual samples of each flavor processed through
complete detector, trigger, and offline simulation. 
Due to the similarity of \ptrel\ for $c$ and $\pi/K$ sources,
within resolution these distributions were indistinguishable.
The $b$ jet signal was extracted on a statistical basis,
through maximum likelihood fits to the observed \ptrel\ distribution.
The normalizations of
Monte Carlo templates representing light ($c$ and $\pi/K$) and heavy 
($b$) quark decay patterns were floated such as to fit the observed \ptrel\ 
spectra in four ranges of transverse energy. 
The extracted fraction of $b$ jets in each $E_T$ range is shown in 
Fig.~\ref{fig:ptrel}.
The systematic uncertainties
were estimated by varying within errors the input distributions to the 
maximum likelihood fits.

The parameterization of the overall fraction of $b$ jets in the inclusive 
sample as a function of jet $E_T$ was obtained from a two-parameter fit
($\chi^2 = 0.96$) to the four bins in $E_T$,
and is shown in Fig.~\ref{fig:ptrel}, 
together with the uncertainty band (6\% relative
uncertainty at 30 GeV, 39\% at 100 GeV). 
The form $(\alpha + \beta/E_T^2)$ was chosen after Monte Carlo fit trials. 
There are partial bin-to-bin correlations in the uncertainty.
The extracted $b$ fraction is the dominant source of
uncertainty in the cross section. The upper reach in $b$ jet $E_T$
is limited by the deteriorating discrimination power of  \ptrel\ 
with increasing $E_T$.

%%%%%%%%%%%%%%%%%%%%%%%%%%%%%%%%%%%%%%%%%%%%%%%%%%%%%%%%%%%%%%%%%%%%%%%%%%%%
%\section*{Tagging Acceptances}
Corrections for acceptance loss stem from 
(i) the tagging threshold for muon \pt\  and range of pseudorapidity,
(ii) the muon-jet association (tag) criterion,
(iii) the undetected lepton energy of the jet, which is added
(statistically) to the hadronic $E_T$ registered in the calorimeter, and
(iv) the restricted pseudorapidity range for jets and muons 
(theory\cite{bib:fmnr} considers $b$ jets with $|\eta|<1$).

The acceptance correction is extracted using the \ISAJET \cite{bib:isajet} 
Monte Carlo simulation, and is defined as the ratio of $b$ jets satisfying 
the analysis conditions to $b$ jets satisfying the cross section definition.
The transition from $E_T$ to $E_T^{b\rm Jet}$ is thus included in
the correction. The simulated $E_T$ and $\eta$ distributions are in 
agreement with the data. 
The overall correction increases from 6\% 
at $E_T=$ 30 GeV to 10\% at 100 GeV.
It is independent of assumed event production rates,
and is affected primarily by models for fragmentation and decay.
Decays are based on the QQ decay package \cite{bib:QQ}
from the CLEO experiment.
\ISAJET\ uses the Peterson fragmentation model \cite{bib:frag},
with a parameter $\epsilon=0.006$ that is varied by $\pm50\%$ to
estimate uncertainty. It is observed that the impact of fragmentation
on the acceptance is mainly due to 
the migration of tagging muons across the minimum-\pt\ boundary. This
uncertainty propagates to the cross section as a 9\% 
effect independent of jet $E_T$.

%%%%%%%%%%%%%%%%%%%%%%%%%%%%%%%%%%%%%%%%%%%%%%%%%%%%%%%%%%%%%%%%%%%%%%%%%%%%
%\section*{b Jet XSecn}
\begin{figure}
\vspace{-1.0cm}
\begin{center}
\mbox{\epsfig{figure=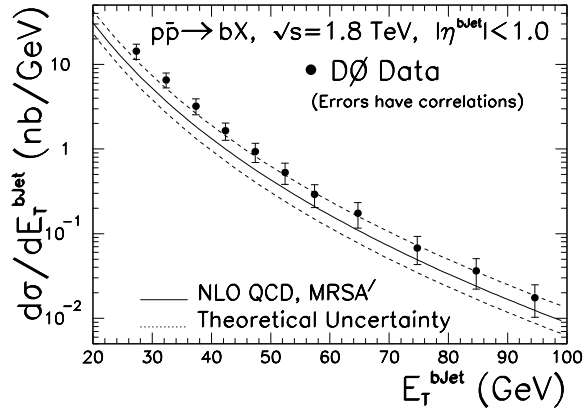,height=6cm,width=8cm}}
\end{center}
\vspace{-0.6cm}
\caption{Differential cross section for $b$ jet production.}
\label{fig:bjxs}
\end{figure}

The resulting $b$ jet cross section for $|\eta|<1$
is shown in Fig.~\ref{fig:bjxs}, together with 
the NLO QCD prediction \cite{bib:fmnr}.
Inputs to this calculation are the renormalization scale
(chosen equal to the factorization scale), 
the $b$ quark pole mass (4.75 GeV/$c^2$),
the parton distribution function (MRSA$^\prime$\cite{bib:mrs}), 
and the parton clustering algorithm. The uncertainty in theory 
is dominated by the QCD renormalization scale dependence, with the 
central value in Fig.~\ref{fig:bjxs}
chosen as \mujot, and varied between $2\mu_0$ (lower curve)
and $\mu_0/2$.
The fact that there is such a strong dependence on scale suggests
that important higher-order terms in the calculation are still missing.

The overall systematic uncertainty in the cross section 
has contributions from
the integrated luminosity (5\%),
trigger (3\%) and offline selection (4\%) efficiencies,
jet $E_T$ scale (15\%--8\%) and resolution effects (7\%--3\%),
tagging acceptance (9\%--12\%),
and the extracted $b$ fraction of jets (6\%--39\%).
Bin-to-bin errors are fully correlated for all sources of uncertainty, 
except for partial correlations in the $b$ fraction of jets.
Cross section values and overall uncertainties, as plotted in 
Fig.~\ref{fig:bjxs}, are listed in Table \ref{tab:bjxs}.

Figure~\ref{fig:bjxs} displays the same general pattern 
of past $b$ production measurements 
\cite{bib:ua1,bib:stich,bib:cdfsgmu,bib:d0sgmu,bib:d0psi,bib:d0corr,bib:samus,bib:cdfmes},
with data lying
above the central values of the prediction, but comparatively less
so in the present case, where general agreement between measurement 
and the upper band of the theoretical uncertainty is observed.

\begin{table}[htb]
\caption{Differential cross section for $b$ jet production.}
\label{tab:bjxs}
\begin{center}
\begin{tabular}{cccc}
$E_{T}^{b\rm Jet}$ & $d\sigma /dE_T^{b\rm Jet}$ 
& \multicolumn{2}{c}{Uncertainties(\%)} \\ %\cline{3-4} 
 (GeV)  &     (nb/GeV)   
        & {\small{Stat.}} 
        &  {\small{Syst.}}      \\ \hline \hline

  27.3  & 14.4     &  1.2 & 20.6     \\ %\hline
  32.3  &  6.57    &  1.6 & 20.1     \\ %\hline
  37.4  &  3.22    &  2.1 & 21.2     \\ %\hline
  42.4  &  1.65    &  2.7 & 23.2     \\ %\hline
  47.4  &  0.932   &  3.4 & 25.6     \\ %\hline
  52.4  &  0.530   &  4.2 & 28.0     \\ %\hline
  57.4  &  0.292   &  5.4 & 30.2     \\ %\hline
  64.7  &  0.174   &  4.6 & 33.2     \\ %\hline
  74.7  &  0.0678  &  6.8 & 36.6     \\ %\hline
  84.7  &  0.0364  &  8.7 & 39.3     \\ %\hline
  94.6  &  0.0176  & 11.7 & 41.5     \\ %\hline
\end{tabular}
\end{center}
\end{table}

To connect the present measurement with previous findings, and in particular 
the apparent difference in normalization with respect to theory, the present 
data sample (muon-tagged jets) can be re-analyzed from a different perspective.
Instead of focusing on the differential $b$ jets cross section, the same data
can be used to reproduce the integrated $b$ quark cross section as a function
of minimum $p_T^b$. Similar measurements are described in Refs.
\cite{bib:ua1,bib:stich,bib:cdfsgmu,bib:d0sgmu} and \cite{bib:d0corr}.
The kinematic relationship between daughter muons ($p_T^\mu$ spectrum) 
and parent $b$ quarks ($p_T^b$ spectrum) is used to extract the $b$ quark 
production cross section, integrated from $p_T^{\rm min}$ to infinity, 
and over the rapidity range $|y^b|<1$. Here
$~p_T^{\rm min}$ is defined as the allowed $p_T^b$ threshold for a given 
$p_T^\mu$ distribution. 
Since the $b$ quark signal fraction increases with muon $p_T$
(decreases with jet $E_T$),
$b$ tagging based on \ptrel\ as a function of $p_T^\mu$ is more precise 
than as a function of jet $E_T$. The model dependence, introduced in the
conversion of the muon \pt\ spectrum to the integrated $b$ quark spectrum,
dominates the uncertainty in this measurement.

\begin{figure}
\vspace{-1cm}
\begin{center}
\mbox{\epsfig{figure=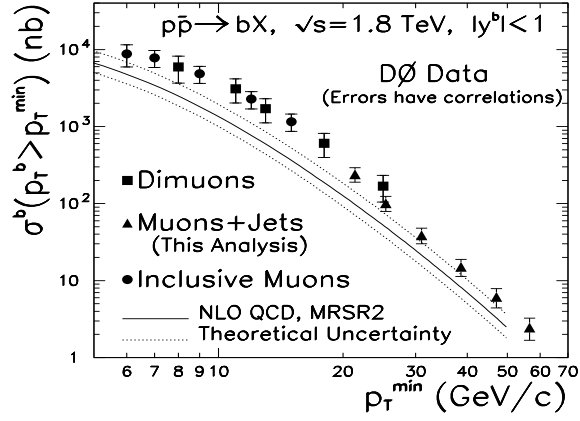,height=6cm,width=8cm}}
\end{center}
\vspace{-0.6cm}
\caption{Integrated cross section for $b$ quark production.}
\label{fig:bqxs}
\end{figure}

The results of this re-analysis are shown in Fig.~\ref{fig:bqxs} 
(triangles) along with the theoretical expectation \cite{bib:nde} 
and the previous \D0 \cite{bib:d0corr} results.
Consistency among all  measurements 
is evident. The present analysis extends significantly 
the $p_T^b$ reach of the previous measurements.
This is due to the requirement of jets in the trigger and offline selection,
as opposed to requiring just muon triggers.
The jet requirement naturally forces the sample into a higher 
region of $p_T^b$. Figure~\ref{fig:bqxs} shows that the
agreement between data and theory improves somewhat at higher $p_T^b$.

%%%%%%%%%%%%%%%%%%%%%%%%%%%%%%%%%%%%%%%%
%\section*{Conclusion}
In conclusion, two distinct measurements of $b$ quark
production in a previously 
unexplored region of larger transverse momenta have been presented, 
and are found to be compatible with the NLO QCD calculation for 
heavy-flavor production. 
Agreement between data and theory, unsatisfactory
for previous measurements, is now observed
to improve at higher transverse momenta.

%%%%%%%%%%%%%%%%%%%%%%%%%%%%%%%%%%%%%%%%
%%\section*{Acknowledgments}
%% Acknowledgement_paragraph.tex
We thank the staffs at Fermilab and at collaborating institutions 
for contributions to this work, and acknowledge support from the 
Department of Energy and National Science Foundation (USA),  
Commissariat  \` a L'Energie Atomique and
CNRS/Institut National de Physique Nucl\'eaire et 
de Physique des Particules (France), 
Ministry for Science and Technology and Ministry for Atomic 
   Energy (Russia),
CAPES and CNPq (Brazil),
Departments of Atomic Energy and Science and Education (India),
Colciencias (Colombia),
CONACyT (Mexico),
Ministry of Education and KOSEF (Korea),
CONICET and UBACyT (Argentina),
A.P. Sloan Foundation,
and the A. von Humboldt Foundation.

%\vspace{-0.4cm}
%%% REFERENCES %%%


\begin{references}
\bibitem{bib:eta}
The pseudorapidity is defined as
$\eta \equiv -{\rm ln}({\rm tan}(\theta/2))$, where $\theta$ is the 
polar angle relative to the proton beam.
\end{references}
\end{document}